\documentclass[aps,onecolumn,showpacs,preprintnumbers,amsmath,amssymb,nofootinbib]{revtex4}
\usepackage{amssymb}
\usepackage{epsfig}
\usepackage{subfigure}
\usepackage{graphicx}
\usepackage{dcolumn}
\usepackage{bm}
\usepackage{pslatex}
\usepackage{slashed}
\usepackage{color}
\begin{document}
\title{$Z_b/Z_b^\prime \to \Upsilon\pi$ and $h_b \pi$ decays in intermediate meson loops model}

\author{Gang Li$^{1}$}
\email{gli@mail.qfnu.edu.cn}
\author{Feng-lan Shao$^1$}
\author{Cheng-Wei Zhao$^{1}$}
\author{Qiang Zhao$^{2,3}$}
\email{zhaoq@ihep.ac.cn}

\affiliation{$^1$Department of Physics, Qufu Normal University, Qufu 273165, China}
\affiliation{$^2$ Institute of High Energy Physics,
       Chinese Academy of Sciences, Beijing 100049, P.R. China}

\affiliation{$^3$ Theoretical Physics Center for Science Facilities,
Chinese Academy of Sciences, Beijing 100049, P.R. China}
\date{\today}

\baselineskip=12pt

\begin{abstract}

With the recent measurement of  $Z_b(10610)$ and $Z_b(10650)\to
B\bar{B}^*+c.c.$ and $B^*\bar{B}^*$, we investigate the transitions
from the $Z_b(10610)$ and $Z_b(10650)$ to bottomonium states with
emission of a pion via intermediate $B \ {B}^*$ meson loops. The
experimental data can be reproduced in this approach with a commonly
accepted range of values for the form factor cutoff parameter
$\alpha$. The $\Upsilon(3S)\pi$ decay channels appear to experience
obvious threshold effects which can be understood by the property of
the loop integrals. By investigating the $\alpha$-dependence of
partial decay widths and ratios between different decay channels, we
show that the intermediate $B \ {B}^*$ meson loops are crucial for
driving the transitions of $Z_b/Z_b'\to \Upsilon(nS)\pi$ with $n =
1, 2, 3$, and $h_b(mP)\pi$ with $m = 1$ and $2$.

\end{abstract}
 \maketitle
\section{Introduction}\label{sec:introduction}

Recently, two charged bottomonium-like structures $Z_b^{\pm}(10610)$
and $Z_b^{\prime \pm}(10650)$ (abbreviated to $Z_b^{\pm}$
and $Z_b^{\prime \pm}$ in the following) were observed by the Belle
Collaboration in the $\pi^{\pm} \Upsilon(nS)$ ($n = 1, 2, 3$) and
$\pi^{\pm} h_b(mP)$ ($m = 1, 2$) invariant mass spectra of
$\Upsilon(5S) \to \Upsilon(nS) \pi^+ \pi^-$ and $h_b(mP) \pi^+
\pi^-$ decays~\cite{Collaboration:2011gja,Belle:2011aa}. The
reported masses and widths of the two resonances are
$M_{Z_b^+} = 10607.2\pm 2.0 $ {\rm MeV}, $\Gamma_{Z_b^+}
= 18.4\pm 2.4$ {\rm MeV} and $M_{Z_b^{\prime +}} = 10652.2\pm
1.5 $ {\rm MeV}, $\Gamma_{Z_b^{\prime +}} = 11.5\pm 2.2$ {\rm
MeV}~\cite{Collaboration:2011gja,Belle:2011aa}. Analyses of the
charged pion angular distributions favor the quantum numbers of the
$Z$-states $I^G(J^P ) = 1^+(1^+)$. Evidence for the charge neutral
partner $Z_b^0$ is found in a Dalitz plot analysis of
$\Upsilon(5S) \to \Upsilon(2S) \pi^0\pi^0$ with $4.9\sigma$
significance by Belle Collaboration~\cite{Adachi:2012im}. Its
measured mass $M_{Z_b^0}=10609_{-6}^{+8} \pm 6$ MeV is also
consistent with that measured in the charged mode. Since $Z_b$'s are
isotriplet states, they need at least four quarks as minimal
constituents, which makes them ideal candidates for exotic hadrons
beyond the conventional $q\bar{q}$ mesons. Note that the decay rates
of $\Upsilon(5S)\to Z_b\pi\to \Upsilon(nS)\pi\pi$ are comparable to
those of $\Upsilon(5S)\to Z_b\pi \to h_b(mP)\pi\pi$. This implies
unusual dynamic mechanisms undergoing the decay process since the
transition to $h_b(mP)$ would require the flip of heavy quark spin
and should be suppressed in the heavy quark mass limit.

Before the observation of $Z_b^+$ and $Z_b^{\prime +}$, the authors predicted the existence of loosely bound
$S$-wave $B{\bar B}^*$ molecular states~\cite{Liu:2008fh,Liu:2008tn}. In Ref.~\cite{Liu:2012vd,Li:2012ss}, the authors predicts the possible existence of $B^{(*)} B^{(*)}$ molecular candidates within one-boson-exchange model. Since the $Z_b^+$ and $Z_b^{\prime +}$ are charged and close to the $B{\bar B}^*$ and $B^* {\bar
B}^*$ thresholds, many studies show that they could be $S$-wave
$B{\bar B}^*$ and $B^* {\bar B}^*$ molecular
states~\cite{Bondar:2011ev,Voloshin:2011qa,Zhang:2011jja,Yang:2011rp,Sun:2011uh,Cleven:2011gp,Mehen:2011yh}.
In Ref.~\cite{Guo:2011gu}, the masses of $S$-wave heavy tetraquarks
$bu {\bar b} {\bar d}$ and $bd {\bar b} {\bar u}$ with $J^P = 1^+$
are extracted  by the chromomagnetic interaction Hamiltonian, which
turn out to be compatible with the corresponding masses of
$Z_b^+$ and $Z_b^{\prime +}$. The QCD sum rule
calculations provide a tetraquark interpretation~\cite{Cui:2011fj}.
Meanwhile, the tetraquark picture is applied to the understanding of
the decays of $Z_b^{\pm}/Z_b^{\prime \pm}\to \pi^\pm
\Upsilon(nS)$ and $\pi^\pm h_b(mP)$~\cite{Ali:2011ug}.


Besides the spectrum study, the production and decay of
$Z_b^+$ and $Z_b^{\prime +}$ are also investigated
extensively. Considering $Z_b^+$ and $Z_b^{\prime +}$
to be $B {\bar B}^*$ and $B^* {\bar B}^*$ molecular states, Voloshin
estimates their production in the radiative decay of
$\Upsilon(5S)$~\cite{Voloshin:2011qa}, and the pion-emission
transitions from $Z_b^+$ and $Z_b^{\prime +}$ to
lighter bottomonia are investigated by
Refs.~\cite{Li:2012uc,Dong:2012hc}. In Ref.~\cite{Cleven:2011gp},
the properties of $Z_b^+$ and $Z_b^{\prime +}$ were
studied in the framework of a nonrelativistic effective field theory
assuming that $Z_b^+$ and $Z_b^{\prime +}$ are the $B
{\bar B}^*$ and $B^* {\bar B}^*$ molecular states.

The intermediate meson loop transitions have been one of the
important nonperturbative transition mechanisms in many processes,
and their impact on the  heavy quarkonium transitions, sometimes
called coupled-channel effects, has been noticed for a long
time~\cite{Lipkin:1986bi,Lipkin:1988tg,Moxhay:1988ri}. By applying
the on-shell approximation, the bottom meson loops were suggested to
play an important role in the $\Upsilon(5S)$ transitions to the
lower $\Upsilon$ states with the emission of two
pions~\cite{Meng:2007tk} or one $\eta$~\cite{Meng:2008bq}. This
mechanism seems to explain many unusual properties that make the
$\Upsilon(5S)$ different from $\Upsilon(4S)$. Similar approach was
also applied to the study of $Z_b$ and $Z_b^\prime$ by Liu {\it et
al.}~\cite{Sun:2011uh}.

In this work, we will investigate the decays of $Z_b \to
\Upsilon(nS)\pi$ and $Z_b \to  h_b(mP)\pi$ via intermediate
$B$-meson loops in an effective Lagrangian approach (ELA) with the
favored quantum numbers $I^G(J^{PC}) = 1^+ (1^{+-})$ for the
$Z_b/Z_b'$. We try to enhance the scenario by quantitative
calculations that the bottomed meson loops are crucial for
explaining the experimental results for $Z_b$ and $Z_b' \to
B\bar{B}^*+c.c.$ and $B^*\bar{B}^*$, and $Z_b$ and $Z_b' \to
\Upsilon(nS)\pi$ and $h_b(mP)\pi$.

The paper is organized as follows. In Sec.~\ref{sec:formula}, we
will introduce the formulae for the ELA.  In Sec.~\ref{sec:results},
the numerical results are presented. The Summary will be given in
Sec.~\ref{sec:summary}.

\begin{figure}[ht]
\centering
\includegraphics[scale=1.0]{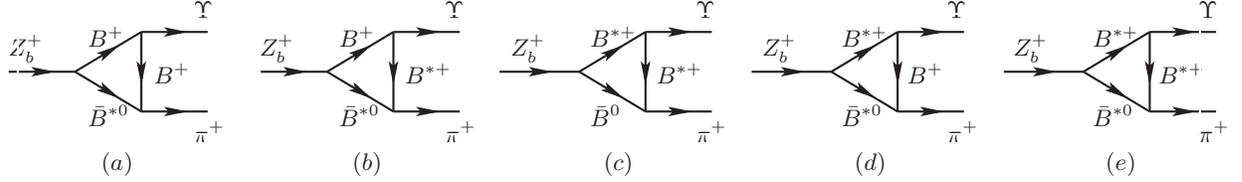}
\caption{The hadron-level diagrams for  $Z_b^+ \to \Upsilon \pi^+$
with $B^{(*)} B^{(*)}$ as the intermediate states. Similar diagrams
for $Z_b^-$ and $Z_b^0$ states decays.}\label{fig:feyn-zb-up}
\end{figure}

\begin{figure}[ht]
\centering
\includegraphics[scale=1.0]{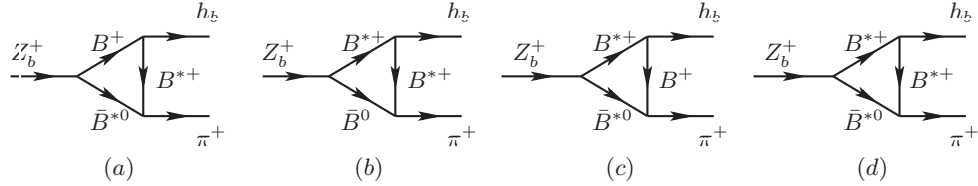}
\caption{The hadron-level diagrams for  $Z_b^+ \to h_b \pi^+$ with
$B^{(*)} B^{(*)}$ as the intermediate states. Similar diagrams for
$Z_b^-$ and $Z_b^0$ states decays.}\label{fig:feyn-zb-hb}
\end{figure}
\section{Transition Amplitude}
\label{sec:formula}

In order to calculate the leading contributions from the bottomed
meson loops, we need the leading order effective Lagrangians for the
couplings. Based on the heavy quark symmetry and chiral
symmetry~\cite{Colangelo:2003sa,Casalbuoni:1996pg}, the relevant
effective Lagrangians used in this work are as follows,
\begin{eqnarray}
\mathcal{L}_{\Upsilon(nS) B^{(*)} B^{(*)}} &=&
ig_{\Upsilon BB} \Upsilon_{\mu} (\partial^\mu B \bar{B}- B
\partial^\mu \bar{B})-g_{\Upsilon B^* B} \varepsilon^{\mu \nu
\alpha \beta}
\partial_{\mu} \Upsilon_{\nu} (\partial_{\alpha} B^*_{\beta} \bar{B}
 + B \partial_{\alpha}
\bar{B}^*_{\beta})\nonumber\\
&&-ig_{\Upsilon B^* B^*} \big\{
\Upsilon^\mu (\partial_{\mu} B^{* \nu} \bar{B}^*_{\nu}
-B^{* \nu} \partial_{\mu}
\bar{B}^*_{\nu})+ (\partial_{\mu} \Upsilon_{\nu} B^{* \nu} -\Upsilon_{\nu}
\partial_{\mu} B^{* \nu}) \bar{B}^{* \mu} +
B^{* \mu}(\Upsilon^\nu \partial_{\mu} \bar{B}^*_{\nu} -
\partial_{\mu} \Upsilon^\nu \bar{B}^*_{\nu})\big\}, \label{eq:h1} \\
\mathcal{L}_{h_b(mP) B^{(*)} B^{(*)}}&=& g_{h_b B^*
B} h_b^\mu ( B \bar{B}^*_{\mu}+ B^*_\mu \bar{B})+ ig_{h_b
B^* B^*} \varepsilon^{\mu \nu \alpha \beta}
\partial_{\mu} h_{b \nu} B^*_{\alpha} \bar{B}^*_{\beta} \ ,\label{eq:h2} \\
{\cal L}_{B^* B^{(*)}\pi} &=& -i g_{B^*B\pi}(B_i \partial_\mu P_{ij} {\bar B}_j^{*\mu} -  B_i^{*\mu} \partial_\mu P_{ij} {\bar B}_j) + \frac {1} {2} g_{B^*B^* \pi} \varepsilon^{\mu\nu\alpha\beta} {B}_{i\mu}^* \partial_\nu P_{ij} {\overleftrightarrow\partial}_\alpha {\bar B}_{j\beta}^*, \label{eq:h3} \\
\mathcal{L}_{Z_b^{(\prime)} B^{(*)} B^{(*)}}&=& g_{Z_b^{(\prime)} B^*
B} Z_b^{(\prime)\mu} ( B \bar{B}^*_{\mu}+ B^*_\mu \bar{B})+ ig_{Z_b^{(\prime)}
B^* B^*} \varepsilon^{\mu \nu \alpha \beta}
\partial_{\mu} Z_{b \nu}^{(\prime)} B^*_{\alpha} \bar{B}^*_{\beta} \ ,\label{eq:h4}
\end{eqnarray}
where
${{B}^{(*)}}=\left(B^{(*)+},B^{(*)0}\right)$ and
${\bar B^{(*)T}}=\left(B^{(*)-},\bar{B}^{(*)0}\right)$ correspond to the
bottom meson isodoublets.

With the experimental data for $BR(Z_b^+ \to B^+ {\bar
B}^{*0} +  {\bar B}^0 B^{*+})=(86.0 \pm 3.6)\%$ and $BR(Z_b^{\prime
+} \to B^{*+} {\bar B}^{*0})= (73.4 \pm 7.0)\%$
from~\cite{Adachi:2012cx}, we obtain $g_{Z_b B^*B} = 13.39$ GeV and
$g_{Z_b^\prime B^*B^*} = 0.32$. The relations
\begin{eqnarray}
g_{Z_b B^*B} = - g_{Z_b B^*B^*}m_{Z_b} \sqrt {\frac {m_B}
{m_{B^*}}},   \quad g_{Z_b^\prime B^*B} = - g_{Z_b^\prime
B^*B^*}m_{Z_b^\prime} \sqrt {\frac {m_B} {m_{B^*}}},
\end{eqnarray}
are applied to extract the couplings for $g_{Z_b B^*B^*}$ and
$g_{Z_b^\prime B^*B}$.

In Eq.~(\ref{eq:h1}), The following couplings are adopted in the
numerical calculations,
\begin{eqnarray}
g_{\Upsilon BB} = 2g_2 \sqrt{m_\Upsilon} m_B \ ,
\quad g_{\Upsilon B^* B} = \frac {g_{\Upsilon BB}} {\sqrt{m_B m_{B^*}}} \ ,
\quad g_{\Upsilon B^* B^*} = g_{\Upsilon B^* B}  \sqrt{\frac {m_{B^*}} {m_B}} m_{B^*} \ ,
\end{eqnarray}
where $g_2= \sqrt{m_\Upsilon}/(2m_B f_\Upsilon)$;  $f_{\Upsilon}$
and $m_{\Upsilon}$ denote the decay constant and mass of
$\Upsilon(nS)$, respectively. The decay  constant $f_{\Upsilon}$ can
be extracted in $\Upsilon(nS)\to e^+e^-$:
\begin{eqnarray}
\Gamma(\Upsilon(nS) \to e^+e^-) = \frac {4\pi\alpha_{\rm EM}^2} {27} \frac {f_{\Upsilon(nS)}^2} {m_{\Upsilon(nS)}},
\end{eqnarray}
where $\alpha_{{\rm EM}} = 1/137$ is the fine-structure constant. By
adopting the mass values in Table~\ref{tab:mass} and data for  the
leptonic decay widths of $\Upsilon(nS)$ states: $\Gamma(\Upsilon(1S)
\to e^+e^-) =1.340 \pm 0.018$ keV, $\Gamma(\Upsilon(2S) \to e^+e^-)
=0.612 \pm 0.011$ keV, $\Gamma(\Upsilon(3S) \to e^+e^-) =0.443 \pm
0.008$ keV~\cite{Beringer:1900zz}, we obtain $f_{\Upsilon(1S)} =
715.2 $ {\rm MeV}, $f_{\Upsilon(2S)} = 497.5 $ {\rm MeV}, and
$f_{\Upsilon(3S)} = 430.2 $ {\rm MeV}.

In addition, the coupling constants in
Eq.~(\ref{eq:h2}) are determined as
\begin{eqnarray}
g_{h_b BB^*} &=& -2g_1 \sqrt{m_{h_b} m_B m_{B^*}} , \ \ g_{h_b
B^* B^*} =2 g_1 \frac{m_{B^*}}{\sqrt{m_{h_b}}},
\end{eqnarray}
with $g_1=-\sqrt{{m_{\chi_{b0}}}/{3}}/{f_{\chi_{b0}}}$, where
$m_{\chi_{b0}}$ and $f_{\chi_{b0}}$ are the mass and decay constant
of $\chi_{b0}(1P)$, respectively~\cite{Colangelo:2002mj}, i.e.
$f_{\chi_{b0}} = 175 \pm 55$ MeV~\cite{Veliev:2010gb},
$f_{\chi_{b0}(2P)}/f_{\chi_{b0}(1P)} =
f_{\Upsilon(2S)}/f_{\Upsilon(1S)}$, and $f_{\chi_{b0}(2P)}=121.6$
MeV.

The coupling constants relevant to the pion interactions in Eq.
(\ref{eq:h4}) are
\begin{eqnarray}
g_{B^* B \pi} = \frac{2
g}{f_\pi} \sqrt{m_B m_{B^*}} \ ,
\quad g_{B^* B^* \pi}=\frac{g_{B^* B \pi}}{\sqrt{m_B m_{B^*}}} \  ,
\end{eqnarray}
where $g=0.44 \pm 0.03_{-0.00}^{+0.01}$~\cite{Becirevic:2009yb} and
$f_{\pi}=132$ MeV are adopted in this work.

The loop transition amplitudes for the transitions in
Figs.~\ref{fig:feyn-zb-up} and \ref{fig:feyn-zb-hb} can be expressed
in a general form in the effective Lagrangian approach as follows:
 \begin{eqnarray}
 M_{fi}=\int \frac {d^4 q_2} {(2\pi)^4} \sum_{D^* \ \mbox{pol.}}
 \frac {T_1T_2T_3} {a_1 a_2 a_3}\prod_i{\cal F}_i(m_i,q_i^2)
 \end{eqnarray}
where $T_i \ (i=1,2,3)$ are the vertex functions; $a_i = q_i^2-m_i^2
\ (i=1,2,3)$ are the denominators of the intermediate meson
propagators. We adopt the form factor, $\prod_i{\cal
F}_i(m_i,q_i^2)$, which is a product of  monopole form factors for
each internal mesons, i.e.
\begin{equation}\label{ELA-form-factor}
\prod_i{\cal F}_i(m_i,q_i^2)\equiv {\cal F}_1(m_1,q_1^2){\cal
F}_2(m_2,q_2^2){\cal F}_3(m_3,q_3^2) \ ,
\end{equation}
with
\begin{equation}{\cal F}_i(m_{i}, q_i^2) \equiv \left(\frac
{\Lambda_i^2-m_{i}^2} {\Lambda_i^2-q_i^2}\right),
\label{ffpara}
\end{equation}
where $\Lambda_i\equiv m_i+\alpha\Lambda_{\rm QCD}$ and the QCD
energy scale $\Lambda_{\rm QCD} = 220$ MeV.
This form factor is supposed to parameterize the non-local effects
of the vertex functions and remove the loop integral divergence.

The explicit transition amplitudes for $Z_b (p_i) \to B^{(*)}(q_1)
B^{(*)} (q_3) [B^{(*)} (q_2)] \to \Upsilon(nS)(p_f) \pi (p_\pi)$ via
those triangle loops are given as follows:
\begin{eqnarray}
M_{BB^* [B]} &=& (i)^3\int \frac {d^4q_2} {(2\pi)^4}[g_{Z_b B^*B}
\varepsilon_{i\mu}] [g_{\Upsilon(nS) BB} \varepsilon_f^{*\rho} (q_1-q_2)_\rho] [g_{B^*B\pi}
p_{\pi\theta}] \nonumber \\
&& \frac {i} {q_1^2-m_1^2}  \frac {i} {q_2^2-m_2^2}  \frac
{i(-g^{\mu\theta} +q_3^\mu q_3^\theta/m_3^2)} {q_3^2-m_3^2} \prod_i{\cal F}_i(m_i,q_i^2) \\
M_{BB^* [B^*]}&=& (i)^3\int \frac {d^4q_2} {(2\pi)^4}[g_{Z_b B^*B}
\varepsilon_{i\mu} ] [g_{\Upsilon(nS) B^*B} \varepsilon_{\rho\sigma \xi\tau}p_f^\rho
\varepsilon_f^{*\sigma} q_2^\xi ] [-g_{B^*B^*\pi}
\varepsilon_{\theta\phi\kappa\lambda}  p_{\pi}^\kappa q_2^\lambda]
\nonumber \\
&& \times \frac {i} {q_1^2-m_1^2}  \frac {i(-g^{\tau\theta}
+q_2^\tau q_2^\theta/m_2^2)} {q_2^2-m_2^2}  \frac {i(-g^{\mu\phi}
+q_3^\mu q_3^\phi/m_3^2)} {q_3^2-m_3^2} \prod_i{\cal F}_i(m_i,q_i^2) \\
M_{B^*B [B^*]} &=& (i)^3\int \frac {d^4q_2} {(2\pi)^4}[g_{Z_b B^*B}
\varepsilon_{i\mu}] [g_{\Upsilon(nS) B^*B^*} (g_{\rho\sigma} g_{\xi\tau} - g_{\rho\tau}
g_{\sigma\xi} + g_{\rho\xi} g_{\sigma\tau}) \varepsilon_f^{*\rho}
(q_1+q_2)^\tau] [-g_{B^*B\pi} p_{\pi\theta}] \nonumber \\
&& \times \frac {i(-g^{\mu\xi} +q_1^\mu q_1^\xi/m_1^2)}
{q_1^2-m_1^2}  \frac {i(-g^{\sigma\theta} +q_2^\sigma
q_2^\theta/m_2^2)} {q_2^2-m_2^2}  \frac {i} {q_3^2-m_3^2} \prod_i{\cal F}_i(m_i,q_i^2) \\
M_{B^*B^* [B]} &=& (i)^3\int \frac {d^4q_2} {(2\pi)^4}[g_{Z_b B^*B^*}
\varepsilon_{\mu\nu\alpha\beta} q_i^\mu\varepsilon_{i}^\nu] [g_{\Upsilon(nS) B^*B}
\varepsilon_{\rho\sigma\xi\tau} p_f^\rho \varepsilon_f^{*\sigma}
q_1^\xi ] [g_{B^*B\pi} p_{\pi\theta}] \nonumber \\
&& \times \frac {i(-g^{\alpha\tau} +q_1^\alpha q_1^\tau/m_1^2)}
{q_1^2-m_1^2}  \frac {i} {q_2^2-m_2^2}  \frac {i(-g^{\beta\theta}
+q_3^\beta q_3^\theta/m_3^2)} {q_3^2-m_3^2} \prod_i{\cal F}_i(m_i,q_i^2) \\
M_{B^*B^* [B^*]} &=& (i)^3\int \frac {d^4q_2} {(2\pi)^4}[g_{Z_b B^*B^*}
\varepsilon_{\mu\nu\alpha\beta} p_i^\mu\varepsilon_{i}^\nu ] [g_{\Upsilon(nS) B^*B^*}
(g_{\rho\sigma} g_{\xi\tau} - g_{\rho\tau} g_{\sigma\xi} +
g_{\rho\xi} g_{\sigma\tau}) \varepsilon_f^{*\rho}  (q_1+q_2)^\tau]
[-g_{B^*B^*\pi}  \varepsilon_{\theta\phi\kappa\lambda} p_{\pi}^\kappa
q_2^\lambda] \nonumber \\ && \times \frac {i(-g^{\alpha\xi}
+q_1^\alpha q_1^\xi/m_1^2)} {q_1^2-m_1^2}  \frac
{i(-g^{\sigma\theta} +q_2^\sigma q_2^\theta/m_2^2)} {q_2^2-m_2^2}
\frac {i(-g^{\beta\phi} +q_3^\beta q_3^\phi/m_3^2)} {q_3^2-m_3^2}
\prod_i{\cal F}_i(m_i,q_i^2) \ .
\end{eqnarray}
Also, the explicit transition amplitudes for $Z_b (p_i) \to B^{(*)}(q_1) B^{(*)} (q_3) [B^{(*)} (q_2)] \to h_b(mP)(p_f) \pi (p_\pi)$
via those triangle loops are given as follows:
\begin{eqnarray}
M_{BB^* [B^*]} &=& (i)^3\int \frac {d^4q_2} {(2\pi)^4}[g_{Z_b B^*B} \varepsilon_{i\mu} ] [g_{h_b(mP) B^*B} \varepsilon_{f\rho}^{*} ] [-g_{B^*B^*\pi}  \varepsilon_{\theta\phi\kappa\lambda} p_{\pi}^\kappa q_2^\lambda] \nonumber \\
&& \times \frac {i} {q_1^2-m_1^2}  \frac {i(-g^{\rho\theta} +q_2^\rho q_2^\theta/m_2^2)} {q_2^2-m_2^2}  \frac {i(-g^{\mu\phi} +q_3^\mu q_3^\phi/m_3^2)} {q_3^2-m_3^2} \prod_i{\cal F}_i(m_i,q_i^2) \\
M_{B^*B [B^*]} &=& (i)^3\int \frac {d^4q_2} {(2\pi)^4}[g_{Z_b B^*B} \varepsilon_{i\mu}] [g_{h_b(mP) B^*B^*} \varepsilon_{\rho\sigma\xi\tau} p_f^\rho \varepsilon_f^{*\sigma} ] [-g_{B^*B\pi}  p_{\pi\theta} ] \nonumber \\
&& \times \frac {i(-g^{\mu\xi} +q_1^\mu q_1^\xi/m_1^2)} {q_1^2-m_1^2}  \frac {i(-g^{\tau\theta} +q_2^\tau q_2^\theta/m_2^2)} {q_2^2-m_2^2}  \frac {i} {q_3^2-m_3^2} \prod_i{\cal F}_i(m_i,q_i^2) \\
M_{B^*B^* [B]} &=& (i)^3\int \frac {d^4q_2} {(2\pi)^4}[g_{Z_b B^*B^*} \varepsilon_{\mu\nu\alpha\beta} p_i^\mu\varepsilon_{0}^\nu ] [g_{h_b(mP) B^*B} \varepsilon_{f\rho}^{*}] [g_{B^*B\pi}  p_{\pi\theta}] \nonumber \\
&& \times \frac {i(-g^{\alpha\rho} +q_1^\alpha q_1^\rho/m_1^2)} {q_1^2-m_1^2}  \frac {i} {q_2^2-m_2^2}  \frac {i(-g^{\beta\theta} +q_3^\beta q_3^\theta/m_3^2)} {q_3^2-m_3^2} \prod_i{\cal F}_i(m_i,q_i^2) \\
M_{B^*B^* [B^*]} &=& (i)^3\int \frac {d^4q_2} {(2\pi)^4}[g_{Z_b B^*B^*} \varepsilon_{\mu\nu\alpha\beta} p_i^\mu\varepsilon_{i}^\nu ] [g_{h_b(mP) B^*B^*} \varepsilon_{\rho\sigma\xi\tau} p_f^\rho \varepsilon_f^{*\sigma} ] [-g_{B^*B^*\pi}   \varepsilon_{\theta\phi\kappa\lambda} p_{\pi}^{\kappa} q_2^\lambda] \nonumber \\
&& \times \frac {i(-g^{\alpha\xi} +q_1^\alpha q_1^\xi/m_1^2)}
{q_1^2-m_1^2}  \frac {i(-g^{\tau\theta} +q_2^\tau q_2^\theta/m_2^2)}
{q_2^2-m_2^2}  \frac {i(-g^{\beta\phi} +q_3^\beta q_3^\phi/m_3^2)}
{q_3^2-m_3^2} \prod_i{\cal F}_i(m_i,q_i^2),
\end{eqnarray}
where $p_i$, $p_f$, $p_\pi$ are the four-vector momenta of the
initial $Z_b$, final state bottomonium and pion, respectively, and
$q_1$, $q_2$, and $q_3$ are the four-vector momenta of the
intermediate bottomed mesons as defined in
Figs.~\ref{fig:feyn-zb-up} and \ref{fig:feyn-zb-hb}.

\section{Results}
\label{sec:results}

\begin{table*}[htb]
\begin{center}
\renewcommand{\arraystretch}{1.3}
\caption{\label{tab:mass}A summary of meson masses adopted in the
calculation.}
\begin{tabular}{ccccccccccccccccc}\hline\hline
States & $\Upsilon(1S)$ & $\Upsilon(2S)$ & $\Upsilon(3S)$ & $h_b(1P)$ & $h_b(2P)$ & $B$ & $B^*$ & $\pi$\\
Mass (MeV)~\cite{Beringer:1900zz}  & $9460$  & $10023$ & $10355$ & $9898$ & $10259$ & $5279$ & $5325$ & $140$ \\
\hline\hline%
\end{tabular}

\end{center}
\end{table*}

\begin{table*}[t]
\begin{center}
\renewcommand{\arraystretch}{1.3}
\caption{\label{tab:loops}List of branching fractions for the
$Z_b^+$ and $Z_b^{\prime +}$ decays. The last column
values are obtained at the average of the central $\alpha$ values
exclude the $\Upsilon(3S)\pi^+$ channels. The experimental values
are taken from~\cite{Adachi:2012cx} as a
reference.}
\begin{tabular}{c|c|c|c|c|c }\hline\hline Initial
states &   Final states  & Exp.  \%  & $\alpha$ range
& ${\bar \alpha}$ value & This work \%\\ \hline
 $Z_b^+(10610)$  &   $\Upsilon(1S)\pi^+$ &   $0.32\pm 0.09$ & $1.47^{+0.18}_{-0.20}$  &  &$0.50$\\
 &   $\Upsilon(2S)\pi^+$ &   $4.38\pm 1.21$  & $1.76^{+0.28}_{-0.29}$  & & $ 4.57$\\
 &   $\Upsilon(3S)\pi^+$ &   $2.15\pm 0.56$  &  $0.51^{+0.09}_{-0.09}$  & $1.81$&$9.53$\\
 &   $h_b(1P)\pi^+$ &  $2.81\pm 1.10$  & $1.76^{+0.24}_{-0.30}$  & &$3.03$\\
 &   $h_b(2P)\pi^+$ &   $4.34\pm 2.07$ & $2.90^{+0.60}_{-0.70}$ & &$1.36$\\ \hline
 $Z_b^{\prime +}(10650)$  &   $\Upsilon(1S)\pi^+$ &   $0.24\pm 0.07$ & $1.23^{+0.16}_{-0.18}$  & &$0.31$\\
 &   $\Upsilon(2S)\pi^+$ &   $2.40\pm 0.63$ & $1.29^{+0.18}_{-0.21}$  & &$2.71$\\
 &   $\Upsilon(3S)\pi^+$ &   $1.64\pm 0.40$ & $ 0.19^{+0.01}_{-0.01}$   & $1.38$ &$5.55$\\
 &   $h_b(1P)\pi^+$ &  $7.43\pm 2.70$ & $1.36^{+0.20}_{-0.24}$  & &$7.72$\\
 &   $h_b(2P)\pi^+$ &   $14.8\pm 6.22$  & $1.62^{+0.38}_{-0.43}$  & & $11.18$\\
\hline\hline%
\end{tabular}

\end{center}
\end{table*}

\begin{table*}[t]
\begin{center}
\renewcommand{\arraystretch}{1.3}
\caption{\label{tab:ratios} The branching ratios of decay rates for
$Z_b^+\to \Upsilon(3S) \pi^+$ and $Z_b^{\prime +}\to
\Upsilon(3S) \pi^+$ with $M_{B^*} = M_B= 5279$ MeV (Scheme-I) and
$M_{B^*} = M_B=5325$ MeV (Scheme-II).}
\begin{tabular}{c | c| c|
c}\hline\hline & channels  &  $Z_b^+(10610)\to \Upsilon(3S)\pi^+$  &
$Z_b^{\prime +}(10650)\to \Upsilon(3S)\pi^+$   \\ \hline Scheme-I
&$\alpha$ & $3.86^{+0.76}_{-0.96}$ & $3.12^{+0.52}_{-0.58}$ \\
\cline{2-4} &
$BR$ (\%)    & $2.15_{-0.57}^{-0.55}$  & $1.64_{-0.40}^{+0.40}$\\
\hline Scheme-II & $\alpha$ & $3.44^{+1.22}_{-0.60}$ &
$3.08^{+0.49}_{-0.50}$ \\ \cline{2-4}
& $BR$ (\%)   & $2.15_{-0.58}^{-0.57}$  & $1.64_{-0.40}^{+0.40}$ \\
\hline\hline%
\end{tabular}

\end{center}
\end{table*}

\begin{figure}[tb]
\begin{center}
\vglue-0mm
\includegraphics[width=0.49\textwidth]{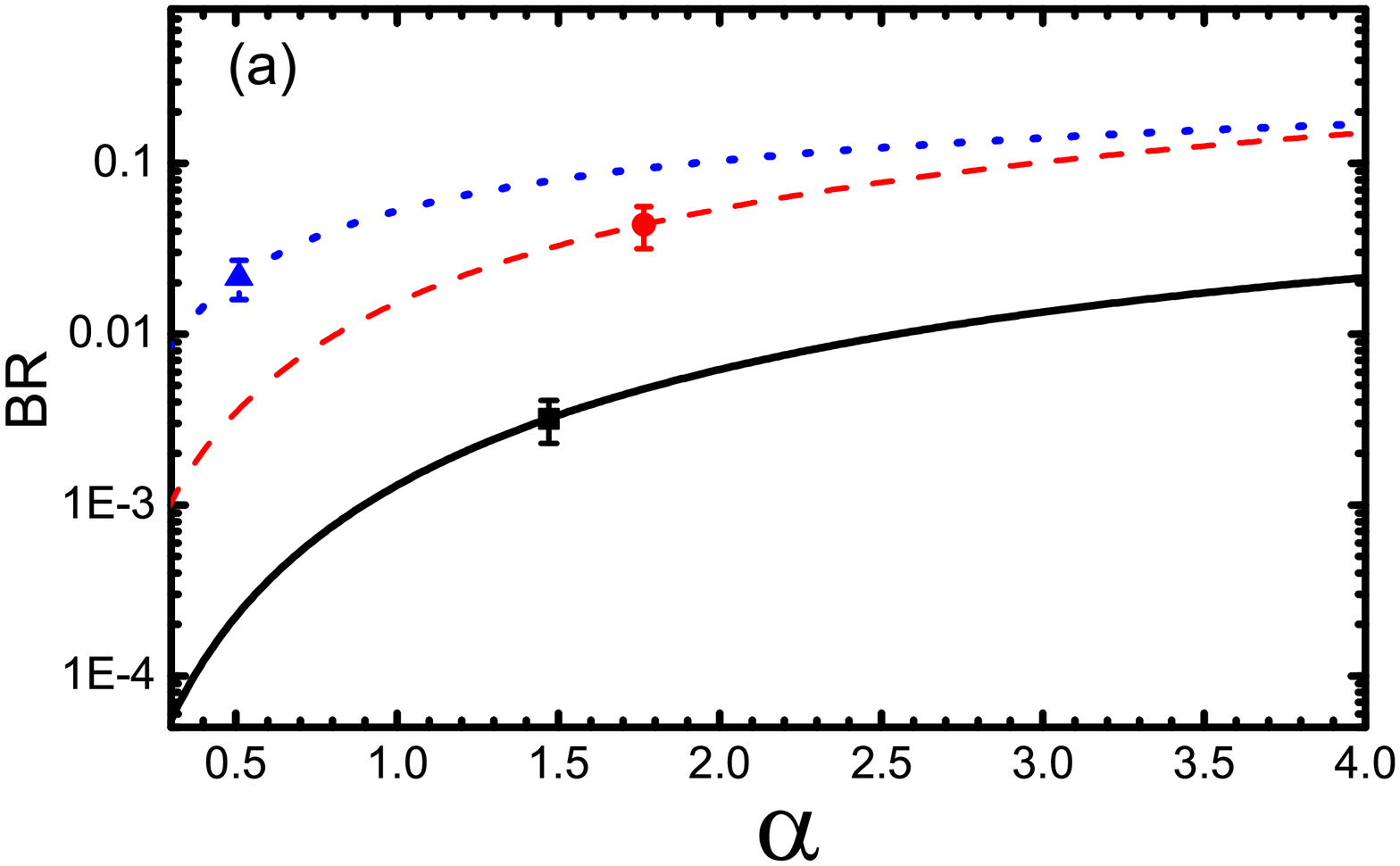}
\includegraphics[width=0.49\textwidth]{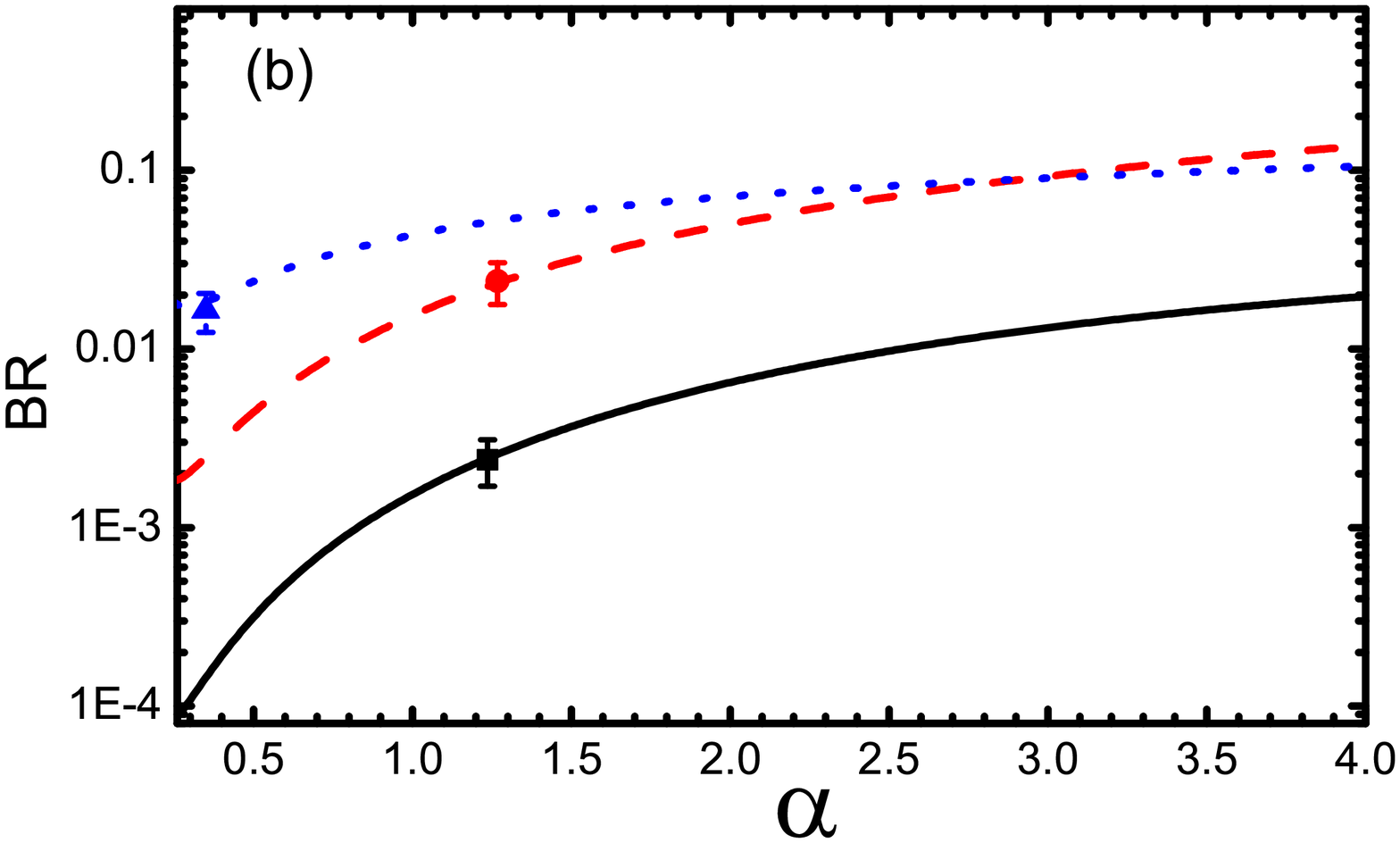}
\caption{ (a) The $\alpha$-dependence of the branching ratios of
$Z_b^+ \to \Upsilon(1S) \pi^+$ (solid line),
$\Upsilon(2S)\pi^+$ (dashed line) and $\Upsilon(3S)\pi^+$ (dotted
line). (b) The $\alpha$-dependence of the branching ratios of
$Z_b^{\prime +} \to \Upsilon(1S) \pi^+$ (solid line),
$\Upsilon(2S)\pi^+$ (dashed line) and $\Upsilon(3S)\pi^+$ (dotted
line). The experimental values
are taken from~\cite{Adachi:2012cx} as a
reference. }\label{fig:alpha_zbuppi1}
\end{center}
\end{figure}

\begin{figure}[tb]
\begin{center}
\vglue-0mm
\includegraphics[width=0.49\textwidth]{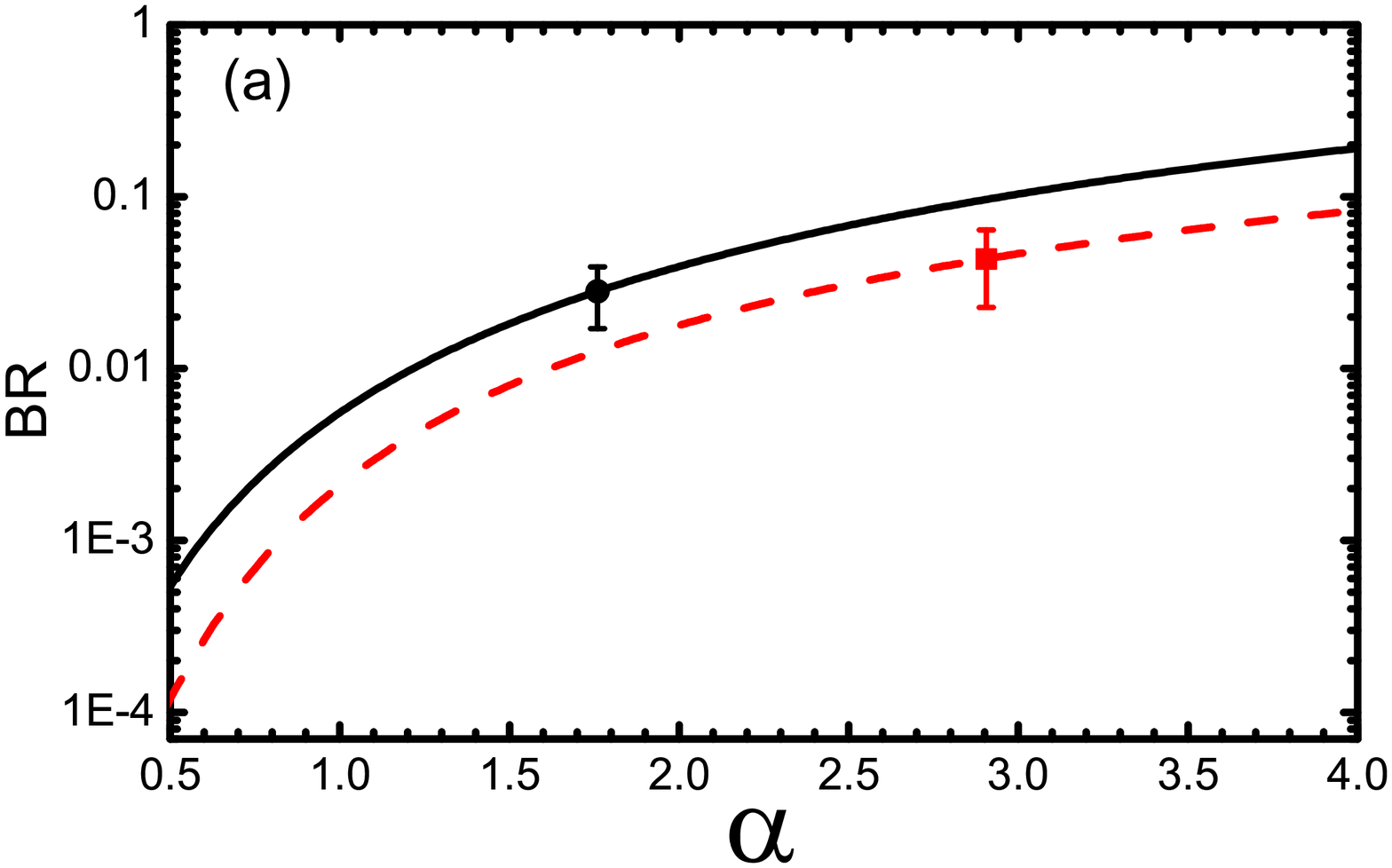}
\includegraphics[width=0.49\textwidth]{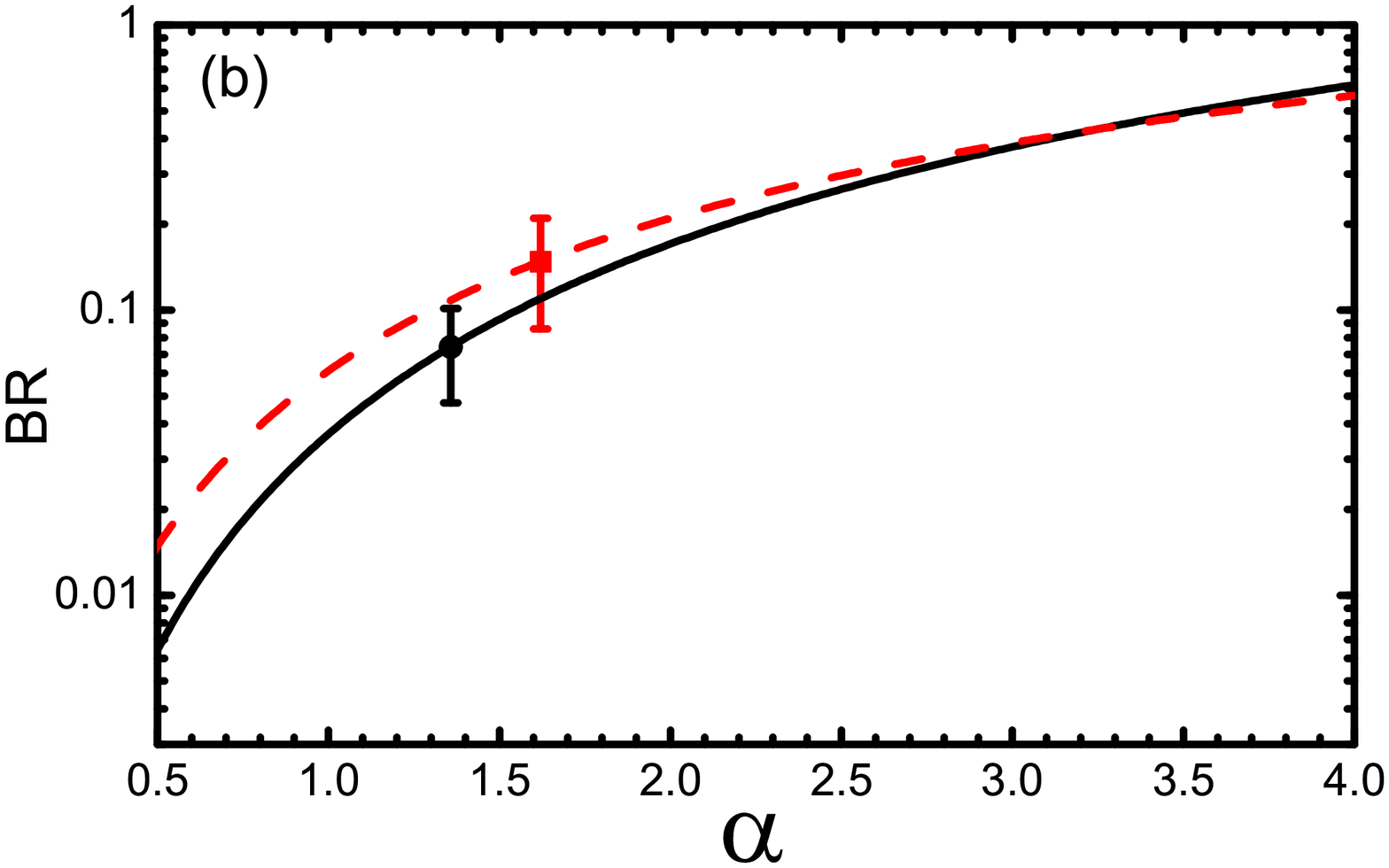}
\caption{(a) The $\alpha$-dependence of the branching ratios of
$Z_b^+ \to h_b(1P) \pi^+$ (solid line) and $h_b(2P) \pi^+$
(dashed line). (b) The $\alpha$-dependence of the branching ratios
of $Z_b^{\prime +} \to h_b(1P) \pi^+$ (solid line) and
$h_b(2P) \pi^+$ (dashed line). The experimental values
are taken from~\cite{Adachi:2012cx} as a
reference.}\label{fig:alpha_zbhbpi1}
\end{center}
\end{figure}

\begin{figure}[tb]
\begin{center}
\vglue-0mm
\includegraphics[width=0.49\textwidth]{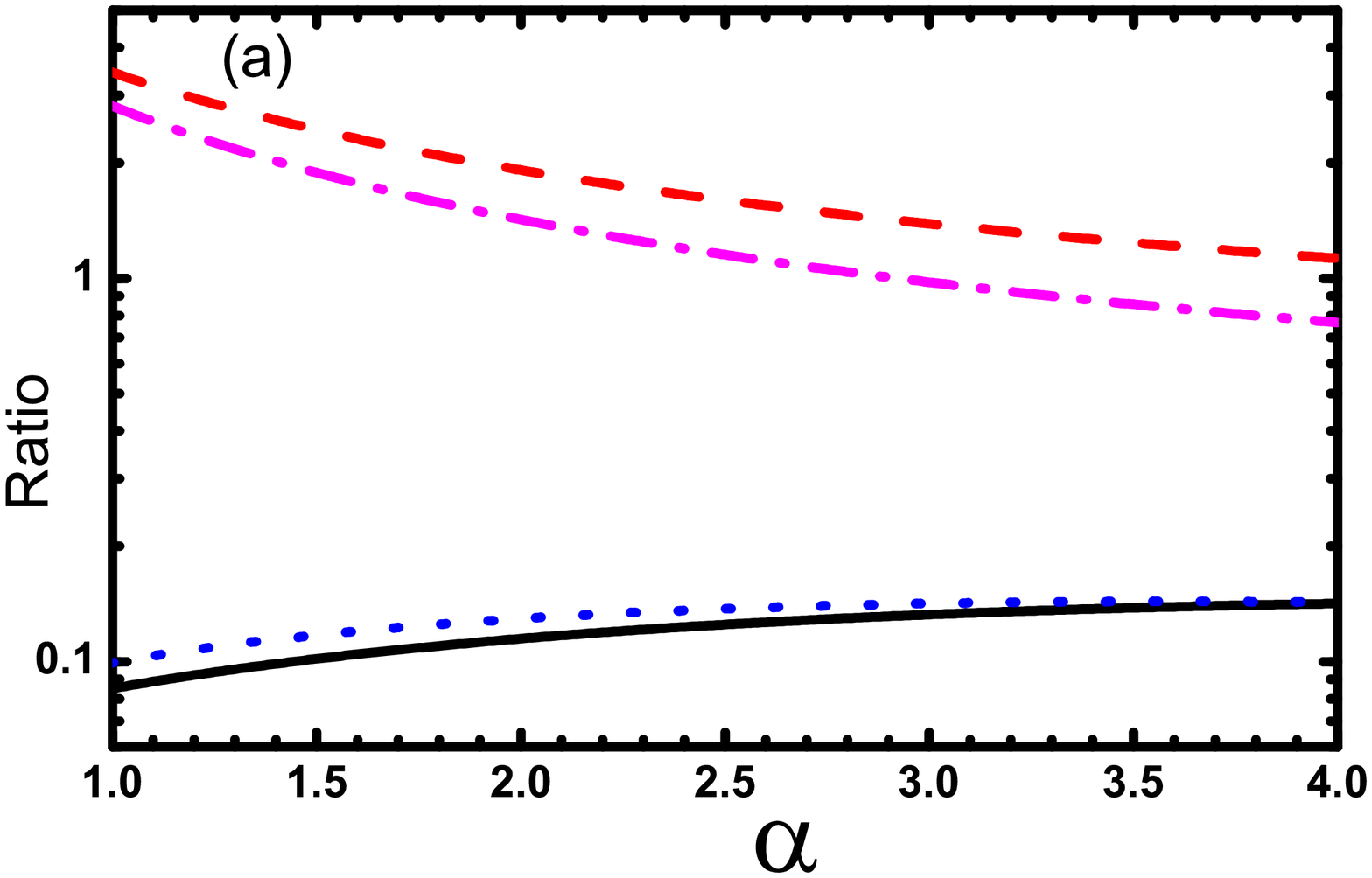}
\includegraphics[width=0.49\textwidth]{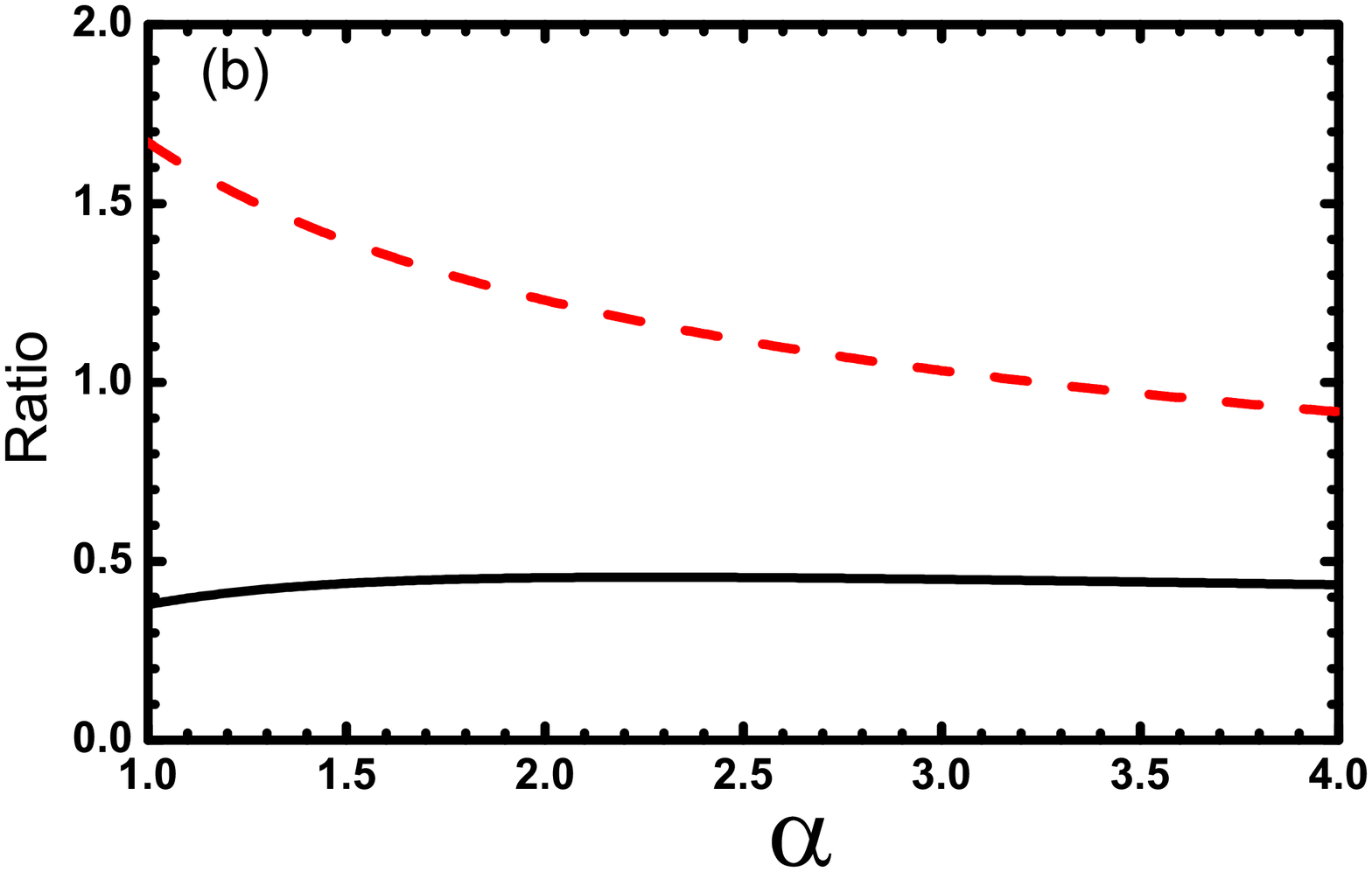}
\caption{(a) The $\alpha$-dependence of the ratios ${\rm
R}_{Z_b}^{12}$ (solid line), ${\rm R}_{Z_b}^{32}$ (dashed line),
${\rm R}_{Z_b^\prime}^{12}$ (dotted line), ${\rm
R}_{Z_b^\prime}^{32}$ (dash-dotted line) defined in
Eq.~(\ref{eq:ratio-1}). (b) The $\alpha$-dependence of the ratios
${\rm r}_{Z_b}$ (solid line), ${\rm r}_{Z_b^\prime}$ (dashed line)
defined in Eq.~(\ref{eq:ratio-2}).}\label{fig:ratio}
\end{center}
\end{figure}

Proceeding to the numerical results, we list the meson masses
involved in the hidden-bottom decays of $Z_b/Z_b^\prime$ in
Table~\ref{tab:mass}. Several points concerning the determination of
the form factor cutoff parameter which would be the only free
parameter in a decay channel, should be clarified. First, we
determine the cutoff parameter $\alpha$ for each channel separately
by the experimental data. As shown in Table~\ref{tab:loops}, it is
possible to find an appropriate range of $\alpha$ values for each
decay channels that can account for the data via the intermediate
bottomed meson loops. Meanwhile, one notices that the $\alpha$
values for $Z_b^+/Z_b^{\prime +}\to \Upsilon(3S)\pi^+$ are much
smaller than other channels which indicates some unusual feature
with this channel. In Table~\ref{tab:loops}, we also list the
$\alpha$ values for each decay channels that can reproduce the
experimental data. An alternative test is that we make an average of
the $\alpha$ values for the $Z_b$ and $Z_b'$ decays separately
without including the $\Upsilon(3S)\pi$ channel, and then check
whether it is possible to describe the experimental data with single
values of $\alpha$ for $Z_b$ and $Z_b'$, respectively.
Interestingly, as shown by the sixth column of
Table~\ref{tab:loops}, with $\alpha=1.81$ and 1.38 for the $Z_b$ and
$Z_b'$ decays, respectively, the data can be reasonably accounted
for except for the $\Upsilon(3S)\pi$ channel.

We also check the $\alpha$-dependence of the decay branching ratios
in order to give a quantitative estimate of the cutoff uncertainties
in the loop integrals. The numerical results are summarized in
Figs.~\ref{fig:alpha_zbuppi1} and \ref{fig:alpha_zbhbpi1} for the
$Z_b$ and $Z_b'$ decays into $\Upsilon(nS)\pi$ and $h_b(mP)\pi$,
respectively.

In Fig.~\ref{fig:alpha_zbuppi1} (a), we plot the $\alpha$ dependence
of the branching ratios of $Z_b^+ \to \Upsilon(1S) \pi^+$ (solid
line), $\Upsilon(2S) \pi^+$ (dashed line), and $\Upsilon(3S) \pi^+$
(dotted line), respectively. A predominant feature is that the
$\alpha$ dependence of the branching ratios are quite stable, which
indicates a reasonable cutoff of the ultraviolet (UV) contributions
by the empirical form factor.  As shown in this figure, at the same
$\alpha$, the intermediate $B$-meson loop effects turn out to be
more important in  $Z_b^+ \to \Upsilon(3S) \pi$ than in $Z_b^+ \to
\Upsilon(1S, \ 2S) \pi^+ $. As a result, a smaller value of $\alpha$
is favored in $Z_b^+ \to \Upsilon(3S) \pi^+$. This is understandable
since the mass of $\Upsilon(3S)$ is closer to the thresholds of
$BB^*$ or $B^*B^*$ than the other two states~\cite{Beringer:1900zz}.
Thus, it gives rise to important threshold effects in $Z_b^+ \to
\Upsilon(3S) \pi^+$.

One notices that the $\alpha$-dependence of the branching ratios for
$Z_b^+/Z_b^{\prime +} \to \Upsilon(3S) \pi^+$ is stabler than those
for $\Upsilon(1S, \ 2S) \pi$. This indicates that the enhanced
branching ratios are not from the off-shell part of the loop
integrals. As we know that the dispersive contributions become
rather model-dependent near threshold, the enhanced (but rather
stable in terms of $\alpha$) branching ratios for $Z_b^+/Z_b^{\prime
+} \to \Upsilon(3S) \pi^+$ suggests that more stringent dynamic
constraints are presumably needed to describe the near-threshold
phenomena where the local quark-hadron duality has been apparently
violated. What makes this process different from e.g. $\psi'\to
h_c\pi^0$ in Ref.~\cite{Guo:2010zk} is that there is no cancelations
between the charged and neutral meson loops. As a consequence, the
subleading terms in Refs.~\cite{Guo:2010zk,Guo:2010ak} become
actually leading contributions. In this sense, a new power counting
in the nonrelativistic effective field theory should be exploited
for the $Z_b/Z_b'$ decays~\cite{q.wang etal}.

Figure~\ref{fig:alpha_zbuppi1} (b) presents the branching ratios of
$Z_b^{\prime +} \to \Upsilon(nS) \pi^+$, and the notation are
the same as Fig.~\ref{fig:alpha_zbuppi1}(a).

The $\alpha$ dependence of the branching ratios of
$Z_b^+/Z_b^{\prime +}\to h_b(1P) \pi^+$  (solid line)
and $h_b(2P) \pi^+$ (dashed line) is presented in
Fig.~\ref{fig:alpha_zbhbpi1}. The experimental data are denoted by
points for corresponding decay channels. The data for
$Z_b^+\to h_b(1P) \pi^+$ and $h_b(2P)\pi^+$ can be
reproduced with $\alpha = 1.76_{-0.30}^{+0.24}$ and
$2.90_{-0.70}^{+0.60}$, respectively. For $Z_b^\prime \to
h_b(1P) \pi^+$ and $h_b(2P) \pi^+$, the values of
$\alpha=1.36^{+0.20}_{-0.24}$ and $1.62^{+0.38}_{-0.43}$ can be
determined by the experimental data. As shown in
Table~\ref{tab:loops}, the decay channels for both $Z_b$ and
$Z_b'\to h_b(1P, \ 2P) \pi$ can be reasonably accounted for
by the averaged values of $\alpha=1.81$ and 1.38, respectively.
Moreover, as shown in Fig.~\ref{fig:alpha_zbhbpi1}, the $\alpha$
dependence turns out to be stable for both $Z_b$ and
$Z_b'$ decays. The stabler behaviors for $Z_b^+$ and
$Z_b^{\prime +}\to h_b(2P) \pi^+$ than $h_b(1P) \pi^+$
indicates the closeness of the $B^*\bar{B}^*$ threshold to the
$h_b(2P)\pi^+$ threshold and the dominance of the meson loop
contributions due to the open threshold effects.


It would be interesting to further clarify the uncertainties arising
from the introduction of form factors by studying the $\alpha$
dependence of the ratios between different partial decay widths. For
the decays of $Z_b^+/Z_b^{\prime +} \to \Upsilon(nS) \pi^+$, we define the
following ratios to the partial decay widths of $Z_b^+/Z_b^{\prime +}\to
\Upsilon(2S)\pi^+$:
\begin{eqnarray}
{\rm R}_{Z_b}^{12}=\frac {\Gamma(Z_b^+ \to \Upsilon(1S) \pi^+)} {\Gamma(Z_b^+ \to \Upsilon(2S) \pi^+)} , \quad
{\rm R}_{Z_b}^{32}=\frac {\Gamma(Z_b^+ \to \Upsilon(3S) \pi^+)} {\Gamma(Z_b^+ \to \Upsilon(2S) \pi^+)} ,  \nonumber \\
{\rm R}_{Z_b^\prime}^{12}=\frac {\Gamma(Z_b^{\prime +} \to \Upsilon(1S) \pi^+)} {\Gamma(Z_b^{\prime +}\to \Upsilon(2S) \pi^+)} , \quad
{\rm R}_{Z_b^\prime}^{32}=\frac {\Gamma(Z_b^{\prime +} \to \Upsilon(3S) \pi^+)} {\Gamma(Z_b^{\prime +}\to \Upsilon(2S) \pi^+)} \, ,\label{eq:ratio-1}
\end{eqnarray}
which are plotted in Fig.~\ref{fig:ratio} (a). The stabilities of
the ratios in terms of $\alpha$ indicate a reasonably controlled
cutoff for each channels by the form factor.

For the decays of $Z_b^+/Z_b^{\prime +}\to h_b(1P, \ 2P) \pi^+$, the
following ratios are defined:
\begin{eqnarray}
{\rm r}_{Z_b}=\frac {\Gamma(Z_b^+ \to h_b(2P) \pi^+)} {\Gamma(Z_b^+ \to
h_b(1P) \pi^+)} , \quad {\rm r}_{Z_b^\prime }=\frac
{\Gamma(Z_b^{\prime +}\to h_b(2P) \pi^+)} {\Gamma(Z_b^{\prime +}\to
h_b(1P) \pi^+)} . \label{eq:ratio-2}
\end{eqnarray}
The $\alpha$ dependence is then plotted in Fig.~\ref{fig:ratio} (b),
which also appears to be highly stable. Since the first coupling
vertices are the same for those decay channels when taking the
ratio, the stability of the ratios suggests that the transitions of
$Z_b/Z_b'\to \Upsilon(nS)\pi$ and $h_b(mP)\pi$ are largely driven by
the open threshold effects via the intermediate $B$ meson loops. In
order to understand this, the following analysis is carried out.
First, one notices that we have adopted the couplings for the $h_b$
and $\Upsilon$ to $B\bar{B}^*$ or $B^*\bar{B}^*$ in the heavy quark
approximation. Since the physical masses for $B$ and $B^*$ are
adopted in the loop integrals, the form factor will introduce
unphysical pole contributions of which the interferences with the
nearby physical poles would lead to model-dependent uncertainties.
By assuming $M_{B^*} = M_B= 5279$ MeV (Scheme-I) and $M_{B^*} =
M_B=5325$ MeV (Scheme-II), namely, by partially restoring the local
quark-hadron duality, we calculate the partial widths of $Z_b^+$ and
$Z_b^{\prime +}\to \Upsilon(3S)\pi$. We expect that the partial
restoration of the local quark-hadron duality will significantly
lower the partial widths since there will be only one physical pole
in the loop and the unphysical one can be easily isolated away from
the physical one. As a result, the inferences caused by the
closeness of the unphysical pole will be reduced~\cite{Guo:2010zk}.
As listed in Table~\ref{tab:ratios} in the heavy quark limit, i.e.
$M_{B^*}=M_B$, the partial widths of $Z_b^+$ and $Z_b^{\prime +}\to
\Upsilon(3S)\pi$ can be reproduced at much larger $\alpha$. This is
a rather direct demonstration of the sensitivity of the meson loop
behaviors when close to open threshold and when the dispersive part
becomes dominant.




In brief, we find it is possible that with the same values of
$\alpha$ for different decay channels, experimental data for the
$Z_b$ and $Z_b'$ hadronic decays can be accounted for in terms of
intermediate $B$ meson loops except for the $\Upsilon(3S)\pi$
channel where the close-to-threshold effect plays an important role.
Recognizing this is helpful for us to understand the experimental
results, and establish the the intermediate $B$ meson loops as the
dominant transition mechanisms for the $Z_b$ and $Z_b'$ decays.



\section{Summary}\label{sec:summary}

In this work, we investigate hidden-bottom decays of the newly
discovered resonances $Z_b^+$ and $Z_b^{\prime +}$ via intermediate
$B$-meson loops. In this calculation, the quantum numbers of the
neutral partners of these two resonances are fixed to be $I^G
(J^{PC}) = 1^+(1^{+-})$, which is currently favored by the
experimental analysis. In the ELA, the present experimental data can
be reproduced with a commonly accepted range of values for the
cutoff parameter $\alpha$ except for the $\Upsilon(3S)\pi$ channel
where the close-to-threshold effect  plays an important role in the
process of $Z_b$ and $Z_b'\to B^*\bar{B}^*(B)\to \Upsilon(3S)\pi$.

Our results show that the $\alpha$ dependence of the branching
ratios are quite stable, which indicate the dominant mechanism
driven by the intermediate meson loops with a fairly well control of
the UV contributions. We also pointed out that the results become
sensitive to the meson loop contributions when the final state mass
threshold are close to the intermediate meson thresholds in our
calculation. Namely, the effects from the unphysical pole introduced
by the form factors would interfere with the nearby physical poles
from the internal propagators and lead to model-dependent
uncertainties. It is also a consequence of the violation of local
quark-hadron duality. Such a phenomenon has been discussed in
Ref.~\cite{Guo:2010ak}. Further experimental and theoretical
studies~\cite{q.wang etal} of the consequences from such an
intermediate meson loop effects would be important for providing
more quantitative information on the structure of $Z_b^+$ and
$Z_b^{\prime +}$.

\section*{Acknowlegements}

Authors thank Q. Wang, F.-K. Guo, C. Hanhart, and U.-G. Mei{\ss}ner
for useful discussions. This work is supported, in part, by the
National Natural Science Foundation of China (Grant Nos. 10947007,
11035006, 11175104, and 11275113), the DFG and NSFC joint project
CRC 110, the Chinese Academy of Sciences (KJCX3-SYW-N2), the
Ministry of Science and Technology of China (2009CB825200), the
Natural Science Foundation of Shandong Province (Grant Nos.
ZR2010AM011 and ZR2011AM006) and the Scientific Research Starting
Foundation of Qufu Normal University.


\begin{thebibliography}{0}
\bibitem{Collaboration:2011gja}
  I.~Adachi [Belle Collaboration],
  arXiv:1105.4583 [hep-ex].

\bibitem{Belle:2011aa}
  A.~Bondar {\it et al.}  [Belle Collaboration],
  Phys.\ Rev.\ Lett.\  {\bf 108}, 122001 (2012)  [arXiv:1110.2251 [hep-ex]].

\bibitem{Adachi:2012im}
  I.~Adachi {\it et al.}  [Belle Collaboration],
  arXiv:1207.4345 [hep-ex].

\bibitem{Liu:2008fh}
  Y.~-R.~Liu, X.~Liu, W.~-Z.~Deng and S.~-L.~Zhu,
  Eur.\ Phys.\ J.\ C {\bf 56}, 63 (2008)  [arXiv:0801.3540 [hep-ph]].

\bibitem{Liu:2008tn}
  X.~Liu, Z.~-G.~Luo, Y.~-R.~Liu and S.~-L.~Zhu,
  Eur.\ Phys.\ J.\ C {\bf 61}, 411 (2009)  [arXiv:0808.0073 [hep-ph]].

\bibitem{Li:2012ss}
  N.~Li, Z.~-F.~Sun, X.~Liu and S.~-L.~Zhu,
  arXiv:1211.5007 [hep-ph].

\bibitem{Liu:2012vd}
  Z.~-W.~Liu, N.~Li and S.~-L.~Zhu,
  arXiv:1211.3578 [hep-ph].

\bibitem{Bondar:2011ev}
  A.~E.~Bondar, A.~Garmash, A.~I.~Milstein, R.~Mizuk and M.~B.~Voloshin,
  Phys.\ Rev.\ D {\bf 84}, 054010 (2011)  [arXiv:1105.4473 [hep-ph]].

\bibitem{Zhang:2011jja}
  J.~-R.~Zhang, M.~Zhong and M.~-Q.~Huang,
  Phys.\ Lett.\ B {\bf 704}, 312 (2011)  [arXiv:1105.5472 [hep-ph]].

\bibitem{Yang:2011rp}
  Y.~Yang, J.~Ping, C.~Deng and H.~-S.~Zong,
  J.\ Phys.\ G {\bf 39}, 105001 (2012)  [arXiv:1105.5935 [hep-ph]].

\bibitem{Sun:2011uh}
  Z.~-F.~Sun, J.~He, X.~Liu, Z.~-G.~Luo and S.~-L.~Zhu,
  Phys.\ Rev.\ D {\bf 84}, 054002 (2011)  [arXiv:1106.2968 [hep-ph]].

\bibitem{Mehen:2011yh}
  T.~Mehen and J.~W.~Powell,
  Phys.\ Rev.\ D {\bf 84}, 114013 (2011)  [arXiv:1109.3479 [hep-ph]].

\bibitem{Voloshin:2011qa}
  M.~B.~Voloshin,
  Phys.\ Rev.\ D {\bf 84}, 031502 (2011)  [arXiv:1105.5829 [hep-ph]].


\bibitem{Cleven:2011gp}
  M.~Cleven, F.~-K.~Guo, C.~Hanhart and U.~-G.~Meissner,
  Eur.\ Phys.\ J.\ A {\bf 47}, 120 (2011)  [arXiv:1107.0254 [hep-ph]].



\bibitem{Guo:2011gu}
  T.~Guo, L.~Cao, M.~-Z.~Zhou and H.~Chen,
  arXiv:1106.2284 [hep-ph].

\bibitem{Cui:2011fj}
  C.~-Y.~Cui, Y.~-L.~Liu and M.~-Q.~Huang,
  Phys.\ Rev.\ D {\bf 85}, 074014 (2012)  [arXiv:1107.1343 [hep-ph]].

\bibitem{Ali:2011ug}
  A.~Ali, C.~Hambrock and W.~Wang,
  Phys.\ Rev.\ D {\bf 85}, 054011 (2012)  [arXiv:1110.1333 [hep-ph]].



\bibitem{Li:2012uc}
  X.~Li and M.~B.~Voloshin,
  arXiv:1207.2425 [hep-ph].


\bibitem{Dong:2012hc}
  Y.~Dong, A.~Faessler, T.~Gutsche and V.~E.~Lyubovitskij,
  arXiv:1203.1894 [hep-ph].

\bibitem{Lipkin:1986bi}
  H.~J.~Lipkin,
  Nucl.\ Phys.\  B {\bf 291}, 720 (1987).

\bibitem{Lipkin:1988tg}
  H.~J.~Lipkin and S.~F.~Tuan,
  Phys.\ Lett.\  B {\bf 206}, 349 (1988).

\bibitem{Moxhay:1988ri}
  P.~Moxhay,
  Phys.\ Rev.\  D {\bf 39}, 3497 (1989).

\bibitem{Meng:2007tk}
  C.~Meng and K.~T.~Chao,
  Phys.\ Rev.\  D {\bf 77}, 074003 (2008)
  [arXiv:0712.3595 [hep-ph]].

\bibitem{Meng:2008bq}
  C.~Meng and K.~T.~Chao,
  Phys.\ Rev.\  D {\bf 78}, 074001 (2008)
  [arXiv:0806.3259 [hep-ph]].

\bibitem{Colangelo:2003sa}
  P.~Colangelo, F.~De Fazio and T.~N.~Pham,
  Phys.\ Rev.\  D {\bf 69}, 054023 (2004)
  [arXiv:hep-ph/0310084].

\bibitem{Casalbuoni:1996pg}
  R.~Casalbuoni, A.~Deandrea, N.~Di Bartolomeo, R.~Gatto, F.~Feruglio and G.~Nardulli,
  Phys.\ Rept.\  {\bf 281}, 145 (1997)
  [arXiv:hep-ph/9605342].

\bibitem{Adachi:2012cx}
  I.~Adachi {\it et al.}  [Belle Collaboration],
  arXiv:1209.6450 [hep-ex].

\bibitem{Beringer:1900zz}
  J.~Beringer {\it et al.}  [Particle Data Group Collaboration],
  Phys.\ Rev.\ D {\bf 86}, 010001 (2012).

\bibitem{Colangelo:2002mj}
  P.~Colangelo, F.~De Fazio and T.~N.~Pham,
  Phys.\ Lett.\ B {\bf 542}, 71 (2002)  [hep-ph/0207061].



\bibitem{Veliev:2010gb}
  E.~V.~Veliev, H.~Sundu, K.~Azizi and M.~Bayar,
  Phys.\ Rev.\ D {\bf 82}, 056012 (2010)  [arXiv:1003.0119 [hep-ph]].


\bibitem{Becirevic:2009yb}
  D.~Becirevic, B.~Blossier, E.~Chang and B.~Haas,
  Phys.\ Lett.\ B {\bf 679}, 231 (2009)  [arXiv:0905.3355 [hep-ph]].

\bibitem{Wu:2011yx}
  J.~-J.~Wu, X.~-H.~Liu, Q.~Zhao and B.~-S.~Zou,
  Phys.\ Rev.\ Lett.\  {\bf 108}, 081803 (2012)  [arXiv:1108.3772 [hep-ph]].

\bibitem{Guo:2010zk}
  F.~-K.~Guo, C.~Hanhart, G.~Li, U.~-G.~Meissner and Q.~Zhao,
  Phys.\ Rev.\ D {\bf 82}, 034025 (2010)  [arXiv:1002.2712 [hep-ph]].


\bibitem{Guo:2010ak}
  F.~-K.~Guo, C.~Hanhart, G.~Li, U.~-G.~Meissner and Q.~Zhao,
  Phys.\ Rev.\ D {\bf 83}, 034013 (2011)  [arXiv:1008.3632 [hep-ph]].


\bibitem{q.wang etal}
Private discussions with Q. Wang {\it et al.} conerning their work
in progress in the framework of nonrelativistic effective field
theory.




\end{thebibliography}
\end{document}